# EMPIRICAL NULL AND FALSE DISCOVERY RATE INFERENCE FOR EXPONENTIAL FAMILIES

By Armin Schwartzman[1]

*Harvard School of Public Health and Dana-Farber Cancer Institute*

In large scale multiple testing, the use of an empirical null distribution rather than the theoretical null distribution can be critical for correct inference. This paper proposes a "mode matching" method for fitting an empirical null when the theoretical null belongs to any exponential family. Based on the central matching method for $z$-scores, mode matching estimates the null density by fitting an appropriate exponential family to the histogram of the test statistics by Poisson regression in a region surrounding the mode. The empirical null estimate is then used to estimate local and tail false discovery rate (FDR) for inference. Delta-method covariance formulas and approximate asymptotic bias formulas are provided, as well as simulation studies of the effect of the tuning parameters of the procedure on the bias-variance trade-off. The standard FDR estimates are found to be biased down at the far tails. Correlation between test statistics is taken into account in the covariance estimates, providing a generalization of Efron's "wing function" for exponential families. Applications with $\chi^2$ statistics are shown in a family-based genome-wide association study from the Framingham Heart Study and an anatomical brain imaging study of dyslexia in children.

**1. Introduction.** In large-scale multiple testing problems, the observed distribution of the test statistics often does not accurately match the theoretical null distribution [Efron et al. (2001), Efron (2004, 2005b)]. In such cases, the use of an empirical null distribution, estimated from the data itself, can be critical for making correct inferences. Previous empirical null methods [Efron (2004, 2007b), Jin and Cai (2007), Efron (2008)] have focused on situations where the theoretical distribution of the test statistics is $N(0, 1)$ or $t$, typically found, for example, in two-group microarray

Received May 2008; revised May 2008.
[1]Supported in part by a William R. and Sara Hart Kimball Stanford Graduate Fellowship.

gene expression studies. Other large-scale multiple testing problems present theoretical null distributions that are not normal or $t$. For instance, $\chi^2$ tests are commonplace in the analysis of genome-wide association studies based on single nucleotide polymorphisms (SNPs) [Van Steen et al. (2005), Kong, Pu and Park (2006)], while multivariate $F$ tests appear in voxel-based analyses of brain imaging studies [Everitt and Bullmore (1999), Schwartzman, Dougherty and Taylor (2005), Lee et al. (2007), Schwartzman et al. (2008b, 2008a)].

This paper extends the scope of the empirical null to distributions that belong to general exponential families, treating the normal and $\chi^2$, as well as their counterparts $t$ and $F$, as special cases. This extension allows the empirical null to be flexibly chosen as a parametric exponential family version of the theoretical null. For example, where the theoretical null $N(0,1)$ may be replaced by an empirical null $N(\mu, \sigma^2)$ with arbitrary mean $\mu$ and variance $\sigma^2$, a theoretical null $\chi^2(\nu_0)$ with fixed $\nu_0$ degrees of freedom may be replaced by a scaled $\chi^2$ density (i.e., gamma) with arbitrary scaling factor $a$ and arbitrary number of degrees of freedom $\nu$ [Schwartzman, Dougherty and Taylor (2008a)].

As a first data example, consider the following family-based study of genome-wide association between genetic variants and obesity based on the Framingham Heart Study (FHS) [Herbert et al. (2006)]. Briefly, genetic markers were obtained by genotyping 1400 probands from the family-plates on an Affymetrix 100K SNP-chip containing 116,204 SNPs. Each SNP was tested for association with four body-mass index measurements at exams 1, 2, 3 and 4 using the multivariate FBAT-GEE statistic [Lange et al. (2003)]. Excluding SNPs for which the number of informative families was less than 20, a total of 95,810 test statistics were generated with theoretical null $\chi^2(4)$. Figure 1(a) shows that the histogram of the test statistics is not as well matched by the theoretical null $\chi^2(4)$ (see zoom-in) as by the empirical null, a scaled $\chi^2$ with 4.27 d.f. and scaling factor 0.95. The mismatch between the histogram and the theoretical null can be seen better in the p-value scale in Figure 1(b). The histogram of p-values according to the empirical null is closer to a uniform distribution than that according to the theoretical null.

A second example where the effect is more dramatic is the brain imaging study analyzed in Schwartzman, Dougherty and Taylor (2008a). In brief, diffusion tensor imaging (DTI) scans were taken of 6 dyslexic and 6 nondyslexic children. After spatial registration, at each of 20,931 voxels a directional test statistic was computed for testing whether the first eigenvector of the mean diffusion tensor has the same 3D spatial orientation in both groups. The scores for each voxel were obtained by a quantile transformation from the theoretical null model $F(2, 20)$ to $\chi^2(2)$. Figure 2(a) shows a histogram of the 20,931 $\chi^2$-scores. The data histogram is not well matched by the theoretical null $\chi^2(2)$ but is better described by the empirical null, a $\chi^2$ with 1.82

EMPIRICAL NULL FOR EXPONENTIAL FAMILIES 3

d.f. This is better seen in Figure 2(b). For p-values that are most likely null, say, higher than 0.1, the theoretical null produces a histogram that can be hardly explained by a uniform distribution. In contrast, the empirical null produces a histogram that is mostly uniform in that range. Moreover, the number of voxels with low p-values (less than 0.05) is higher according to the empirical null, indicating a gain in statistical power. Schwartzman et al. (2008b) show other examples of voxel-based analyses in brain imaging with normal and $\chi^2$ statistics where the empirical null is necessary for correct inference.

The proposed method for fitting the empirical null, which I call 'mode matching', is a generalization of the central matching method for $z$-scores [Efron et al. (2001), Efron (2004, 2007b)]. Mode matching consists of fitting the empirical null to a region of the histogram of the test statistics surrounding the mode, which for the normal distribution coincides with matching the center. Mode matching is presented here with a one-step approach, fitting the empirical null to the histogram directly by Poisson regression. This contrasts with the two-step scheme of Efron (2007b), where a nonparametric density is first fitted to the histogram by Poisson regression and then the empirical null is fitted to the nonparametric density estimate by least squares. The one-step fit simplifies the theoretical analysis of bias and variance and avoids the need to tune additional parameters for nonparametric density estimation. But, most importantly, it highlights why mode matching is effective for exponential families: for these the log-link function of the Poisson regression becomes linear in the regression parameters.

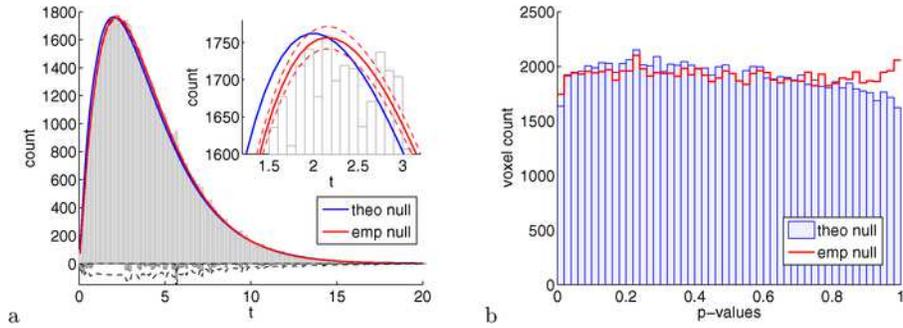

FIG. 1. *SNP example:* (a) *Histogram of the test statistics (light gray). Superimposed densities are the theoretical null $\chi^2(4)$ (dashed) and the empirical null (solid) with pointwise standard 95% CIs. The histogram of the estimated alternative component and corresponding upper standard CI are shown in inverted scale. Inlet plot is a zoom-in.* (b) *Histogram of p-values according to the theoretical null (light gray) and the empirical null (black).*

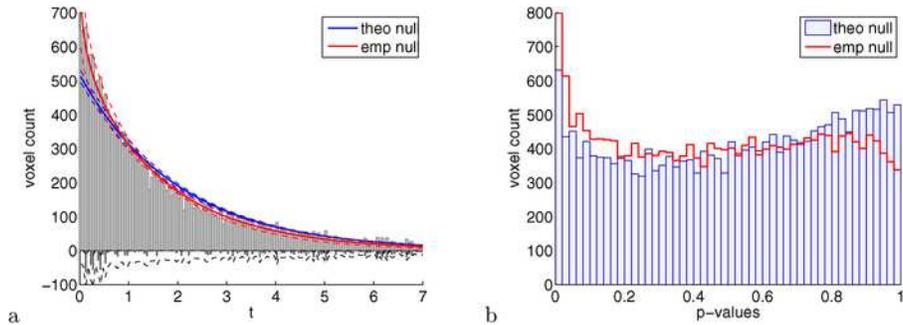

FIG. 2. *DTI example:* (a) *Histogram of the $\chi^2$-scores (light gray). Superimposed densities are the theoretical null $\chi^2(2)$ (dashed) and the empirical null (solid) with pointwise standard 95% CIs. The histogram of the estimated alternative component and corresponding upper standard CI are shown in inverted scale.* (b) *Histogram of p-values according to the theoretical null (light gray) and the empirical null (black).*

The empirical null may be used with any multiple testing procedure. Nonetheless, mode matching is particularly suited for estimating the false discovery rate (FDR), a commonly used error measure in multiple testing problems [Benjamini and Hochberg (1995), Genovese and Wasserman (2004), Storey, Taylor and Siegmund (2004)]. Below I present formulas for calculating the local and tail FDR estimates and show that, as with central matching [Efron (2005b, 2007b)], these estimates follow easily from mode matching calculations for general exponential families.

Delta method covariance formulas are derived for both the empirical null and FDR estimates. It is shown that these formulas produce variance estimates similar to those obtained by the bootstrap when the test statistics are independent. Permutations are used to respect the correlation between test statistics when they are not independent.

Further, approximate formulas are derived for the bias of both the empirical null and FDR estimates. It is shown that the bias in the empirical null is driven mainly by the likelihood ratio between the alternative and null distributions. Simulations are used to inform the choice of the two tuning parameters of mode matching (histogram bin width and fitting interval) in terms of the bias-variance trade-off. For example, in agreement with Efron (2007b), it is found that in the normal case mode matching is fairly insensitive to the choice of bin width, but in the $\chi^2$ case the choice of bin width is affected by the curvature of the density, which sharply increases when the number of degrees of freedom is less than 2. The fitting interval plays a more important role, as it controls the bias introduced by the alternative distribution. In terms of FDR estimation, the bias formulas reveal that both the local and tail FDR estimates can be deceptively biased down for very high thresholds (low p-values), where the number of observed test statistics is low. I argue that this effect should be carefully taken into account when making inferences in real data sets.

The effect of dependence in the covariance of the empirical null and FDR estimates is explained in terms of the empirical distribution of pairwise correlation between test statistics in a way similar to Efron (2007a). I show that Efron's enigmatic "wing function" is a special case of the large family of Lancaster polynomials of bivariate exponential families, which reduces to the Hermite polynomials in the normal case and to the Laguerre polynomials in the $\chi^2$ case.

Mode matching is both computationally efficient and easy to implement because it is based on Poisson regression, for which software is widely available. The analysis is demonstrated in both the DTI and SNP examples introduced above. The SNP example demonstrates the bias, while the DTI example demonstrates the effect of correlation. While both examples have $\chi^2$ null distributions, I emphasize that the methodology is designed for general exponential families. Specific procedures and simulation results are shown for both the normal and $\chi^2$ cases.

## 2. Mode matching for exponential families.

2.1. *Setup.* Let $T_1, \ldots, T_N$ be a large collection of $N$ test statistics. The two-class mixture model [Efron et al. (2001), Storey (2003), Efron (2004, 2007b) Sun and Cai (2007)]

$$(1) \qquad f(t) = p_0 f_0(t) + (1 - p_0) f_A(t)$$

specifies that a fixed fraction $p_0$ of the test statistics behave according to a common null distribution with density $f_0(t)$. The other test statistics behave according to alternative densities whose mixture is $f_A(t)$. The null density $f_0(t)$ is assumed unimodal. The zero assumption, needed for identifiability of the model, is loosely defined by Efron as the condition that most of the probability mass near the mode of $f(t)$ is due to the null term $p_0 f_0(t)$, for example, $p_0 > 0.9$ (the effect of overlap between the null and alternative components is discussed in Section 4). The objective of the empirical null methodology is to estimate $p_0$ and $f_0$ from $T_1, \ldots, T_N$.

Mode matching begins by summarizing the data into a vector of histogram counts $\boldsymbol{y} = (y_1, \ldots, y_K)'$ with $y_k = \sum_{i=1}^N \mathbf{1}\{T_i \in B_k\}$, $k = 1, \ldots, K$, for $K$ bins $B_k$ centered at $\boldsymbol{t} = (t_1, \ldots, t_K)'$. For simplicity, I assume all bins have the same width $\Delta$, although this is not crucial. If the test statistics are independent, then, given $N$, the counts $\boldsymbol{y}$ follow a multinomial distribution with probabilities $\boldsymbol{\pi} = (\pi_1, \ldots, \pi_K)'$, $\pi_k = P(T_i \in B_k)$. By the Taylor expansion around $t_k$,

$$(2) \qquad \pi_k = \int_{B_k} f(t)\,dt = \Delta f(t_k) + \frac{\Delta^3}{24} f''(t_k) + \cdots \approx \Delta f(t_k).$$

The approximation is valid if the bin width $\Delta$ is small and the marginal density $f(t)$ is smooth (the effect of curvature is discussed in Section 4). Thus, for large $N$, the scaled histogram

$$\hat{f}(\boldsymbol{t}) = \frac{\boldsymbol{y}}{N\Delta} \tag{3}$$

is a nearly unbiased estimate of $f(\boldsymbol{t})$ at the bin centers $\boldsymbol{t}$.

The next step is to choose a closed interval $S_0$ where the zero assumption may hold. $S_0$ is the union of $K_0 < K$ consecutive bins containing the mode of $f(t)$. For example, for a two-sided test with theoretical null $N(0,1)$, $S_0$ may be of the form $S_0 = [t_{\min}, t_{\max}]$, while for a one-sided test with theoretical null $\chi^2$, $S_0$ may be of the form $S_0 = [0, t_{\max}]$. Within $S_0$, the zero assumption makes (3) an estimate of the scaled null $p_0 f_0(t)$ in (1), with additional bias $(1-p_0)f_A(t)$.

Suppose $f_0(t)$ is a parametric density. Instead of maximizing the multinomial likelihood given $\boldsymbol{y}$, mode matching uses, almost equivalently, Poisson regression. The idea, also called Lindsey's method [Efron and Tibshirani (1996), Efron (2007b)], is to consider the number of tests $N$ as a Poisson variable $N \sim Po(\gamma)$. If the test statistics are independent, then the histogram counts become independent Poisson variables $y_k \sim Po(\lambda_k)$ with $\lambda_k = \gamma \pi_k$. If $N$ is large, this is essentially the same as the usual Poisson approximation to the multinomial. Using (2), we have $\lambda_k = \gamma \pi_k \approx \gamma \Delta f(t_k)$. Thus, within $S_0$, the zero assumption leads to the general Poisson regression model $y_k \sim Po(\lambda_k)$ with

$$\lambda_k \approx \gamma \Delta p_0 f_0(t_k), \qquad t_k \in S_0, \tag{4}$$

where $\gamma$ is replaced by its MLE, the observed count $N$.

2.2. *Exponential families.* Since the link function for Poisson regression is logarithmic, the precise parametric form of $f_0(t)$ needed to make $\log(\lambda_k)$ in (4) linear in the parameters is an exponential family. Let

$$f_0(t) = g_0(t) \exp(\boldsymbol{x}(t)'\boldsymbol{\eta} - \psi(\boldsymbol{\eta})), \tag{5}$$

where $g_0(t)$ is the carrier density, $\boldsymbol{\eta}$ is the vector of canonical parameters, $\boldsymbol{x}(t)$ is the sufficient vector and $\psi(\boldsymbol{\eta})$ is the cumulant generating function. Replacing in (4) gives the linear Poisson regression model $y_k \sim Po(\lambda_k)$ with

$$\log(\lambda_k) = \boldsymbol{x}(t_k)'\boldsymbol{\eta} + C + h_k, \tag{6}$$

where the entries of $\boldsymbol{x}(t_k)$ play the role of predictors,

$$C = C(\boldsymbol{\eta}) = \log p_0 - \psi(\boldsymbol{\eta}) \tag{7}$$

is a constant intercept, and $h_k = \log(N\Delta g_0(t_k))$ is an offset. It is convenient to write model (6) in vector form as

$$\log(\boldsymbol{\lambda}) = \boldsymbol{X}\boldsymbol{\eta}^+ + \boldsymbol{h}, \tag{8}$$

where $\boldsymbol{\lambda} = (\lambda_1, \ldots, \lambda_K)'$, $\boldsymbol{\eta}^+ = (C, \boldsymbol{\eta}')'$ is the augmented parameter vector, the design matrix $\boldsymbol{X}$ has rows $(1, \boldsymbol{x}(t_k)')$ for $k = 1, \ldots, K$, and $\boldsymbol{h} = (h_1, \ldots, h_K)'$. The fit is restricted to the interval $S_0$ by providing the Poisson regression algorithm with an external set of weights $\boldsymbol{w} = (w_1, \ldots, w_K)'$, where $w_k$ is equal to 1 or 0 according to whether $t_k$ is in $S_0$ or not. For later use, define the diagonal matrix $\boldsymbol{W}$ with diagonal equal to $\boldsymbol{w}$ (not to be confused with the weighting matrix used internally in the iterative solving of the Poisson regression).

Solving (8) gives estimates $\hat{\boldsymbol{\eta}}^+ = (\hat{C}, \hat{\boldsymbol{\eta}})'$, which include the empirical null parameter estimates $\hat{\boldsymbol{\eta}}$. From these, an estimate of the null probability $p_0$ is also obtained using (7) as $\hat{p}_0 = \exp(\hat{C} + \psi(\hat{\boldsymbol{\eta}}))$. Notice that $\hat{p}_0$ is not constrained to be less than or equal to 1. The predicted histogram counts $\hat{\boldsymbol{\lambda}} = N\Delta \hat{f}_0(\boldsymbol{t}) = \hat{\boldsymbol{y}} = (\hat{y}_1, \ldots, \hat{y}_K)'$ corresponding to the empirical null for all bins (not just within $S_0$) are

$$(9) \qquad \hat{\boldsymbol{y}} = \exp(\boldsymbol{X}\hat{\boldsymbol{\eta}}^+ + \boldsymbol{h}).$$

As a result, the predicted histogram counts corresponding to the alternative component in (1) are

$$(10) \qquad N\Delta(1 - \hat{p}_0)\hat{f}_A(\boldsymbol{t}) = N\Delta(\hat{f}(\boldsymbol{t}) - \hat{p}_0 \hat{f}_0(\boldsymbol{t})) = \boldsymbol{y} - \hat{\boldsymbol{y}}.$$

Empirical null densities are more naturally specified using the usual parameters of the distribution rather than the canonical ones. When the theoretical null is $N(0,1)$, the empirical null is $N(\mu, \sigma^2)$ with $\boldsymbol{\theta} = (\mu, \sigma^2)'$ [Efron (2004, 2007b)] [$t$-statistics are handled by a quantile transformation to $N(0,1)$]. When the theoretical null is $\chi^2$ with $\nu_0$ d.f., an appropriate empirical null is a scaled $\chi^2$ with $\nu$ d.f. and scaling factor $a$, denoted $a\chi^2(\nu)$, with density

$$(11) \qquad f_0(t) = \frac{1}{(2a)^{\nu/2}\Gamma(\nu/2)} e^{-t/(2a)} t^{\nu/2-1},$$

where $\boldsymbol{\theta} = (a, \nu)'$ [Schwartzman, Dougherty and Taylor (2008a)]. This is the same as a gamma density with shape parameter $\nu/2$ and scaling parameter $2a$, but using the $\chi^2$ notation helps keep the connection to the theoretical null. $F$-statistics are handled by a quantile transformation to $\chi^2$ with the same numerator number of degrees of freedom.

Let $\boldsymbol{\theta} = \boldsymbol{\theta}(\boldsymbol{\eta})$ denote the vector of usual parameters as in the normal and $\chi^2$ examples above. Let $\boldsymbol{\theta}^+ = (\log p_0, \boldsymbol{\theta}')'$ be the augmented parameter vector. The MLE of $\boldsymbol{\theta}^+$ is $\hat{\boldsymbol{\theta}}^+ = (\log \hat{p}_0, \boldsymbol{\theta}(\hat{\boldsymbol{\eta}})')'$. The derivation of these parameter estimates from the canonical parameter estimates for both the normal and $\chi^2$ cases is worked out in Appendix A.

Other distributions are treated in a similar way. For p-values, whose theoretical null is uniform, the empirical null may be a beta distribution with fitting interval $S_0 = [t_{\min}, 1]$. If the theoretical null is a discrete exponential family (e.g., binomial, Poisson, negative binomial), the mode matching procedure is the same as above except that the bins width $\Delta = 1$ is automatically set by the discrete nature of the distribution, making equations (2) and (4) exact rather than approximate.

2.3. *Exponential subfamilies.* In some cases, one may want to adjust only some of the parameters in (5) and leave the others fixed as prescribed by the theoretical null. For instance, the microarray analysis examples in Efron (2007b) suggest the empirical null $N(0, \sigma^2)$, while in some fMRI studies involving $z$-scores, an appropriate empirical null may be $N(\mu, 1)$ [Ghahremani and Taylor (2005)]. If fixing some parameters results in another lower dimensional exponential family, then the procedure is similar to the one above after the canonical parameters have been redefined. Let $\boldsymbol{\eta}^+ = (C, \boldsymbol{\eta}_1', \boldsymbol{\eta}_2')'$, where $\boldsymbol{\eta}_1$ is the vector of canonical parameters to be estimated and $\boldsymbol{\eta}_2$ is the vector of parameters whose values are fixed. Let $\boldsymbol{X} = (\boldsymbol{1}_K, \boldsymbol{X}_1, \boldsymbol{X}_2)$ be the corresponding split of the design matrix, where $\boldsymbol{1}_K$ indicates a column of $K$ ones. The regression equation (8) becomes

$$\log(\boldsymbol{\lambda}) = \boldsymbol{1}_K C + \boldsymbol{X}_1 \boldsymbol{\eta}_1 + (\boldsymbol{X}_2 \boldsymbol{\eta}_2 + \boldsymbol{h}) \tag{12}$$

and is solved as before, except that the fixed term $\boldsymbol{X}_2 \boldsymbol{\eta}_2$ is absorbed into the offset in parenthesis. The specific exponential subfamilies of the normal and $\chi^2$ cases are worked out in detail in Appendix A.

The simplest restricted case is where one believes the theoretical null and no adjustment of parameters is necessary, except for $p_0$ [Efron (2004)]. In that case, only the intercept $C$ needs to be estimated in (12), treating all the other terms as offset. The estimate of $p_0$ is then given by $\hat{p}_0 = \exp(\hat{C} + \psi(\boldsymbol{\eta}))$. Notice that, for the regression (12) to remain linear, $p_0$ cannot be fixed a priori.

2.4. *Covariance estimates.* Covariance estimates for the empirical null parameter estimates $\hat{\boldsymbol{\eta}}^+$ can be obtained by the delta method in a way similar to Efron (2005b). For this we first need an estimate of the covariance of $\boldsymbol{y}$. As noted by Efron and Tibshirani (1996), there are two such estimates. The Poisson regression regards the observations $y_k$ as independent, so its influence function is determined by the covariance estimate $\widehat{\text{cov}}(\boldsymbol{y}) = \hat{\boldsymbol{V}} = \text{Diag}(\hat{\boldsymbol{y}})$, a $K \times K$ diagonal matrix with diagonal entries $\hat{y}_k$. On the other hand, the true covariance of $\boldsymbol{y}$ depends on the dependence structure of the test statistics.

Suppose first the test statistics are independent. Then conditional on $N$, the $y_k$ are multinomial, for which an appropriate covariance estimate is

$$\hat{\boldsymbol{V}}_N = \text{Diag}(\hat{\boldsymbol{y}}) - \hat{\boldsymbol{y}}\hat{\boldsymbol{y}}'/N. \tag{13}$$

PROPOSITION 1. *Let $\dot{\psi}(\hat{\boldsymbol{\eta}})$ and $\dot{\boldsymbol{\theta}}(\hat{\boldsymbol{\eta}})$ denote the derivatives of $\psi$ and $\boldsymbol{\theta}$ with respect to $\boldsymbol{\eta}$ evaluated at $\hat{\boldsymbol{\eta}}$. The delta method covariance estimates of $\hat{\boldsymbol{\eta}}^+$ and $\hat{\boldsymbol{\theta}}^+$ are respectively*

$$(14) \qquad \widehat{\mathrm{cov}}(\hat{\boldsymbol{\eta}}^+) = (\boldsymbol{X}'\boldsymbol{W}\hat{\boldsymbol{V}}\boldsymbol{X})^{-1}\boldsymbol{X}'\boldsymbol{W}\hat{\boldsymbol{V}}_N\boldsymbol{W}\boldsymbol{X}(\boldsymbol{X}'\boldsymbol{W}\hat{\boldsymbol{V}}\boldsymbol{X})^{-1}$$

$$(15) \qquad \widehat{\mathrm{cov}}(\hat{\boldsymbol{\theta}}^+) = \hat{\boldsymbol{D}}\,\widehat{\mathrm{cov}}(\hat{\boldsymbol{\eta}}^+)\hat{\boldsymbol{D}}', \qquad \hat{\boldsymbol{D}} = \begin{pmatrix} 1 & \dot{\psi}(\hat{\boldsymbol{\eta}})' \\ 0 & \dot{\boldsymbol{\theta}}(\hat{\boldsymbol{\eta}})' \end{pmatrix}.$$

PROPOSITION 2. *The delta method covariance estimate of the empirical null fits (9) and the empirical alternative component (10) are respectively $\widehat{\mathrm{cov}}(\hat{\boldsymbol{y}}) = (\hat{\boldsymbol{V}}\boldsymbol{D}_y)\hat{\boldsymbol{V}}_N(\hat{\boldsymbol{V}}\boldsymbol{D}_y)'$ and $\widehat{\mathrm{cov}}(\boldsymbol{y}-\hat{\boldsymbol{y}}) = (\boldsymbol{I}-\hat{\boldsymbol{V}}\boldsymbol{D}_y)\hat{\boldsymbol{V}}_N(\boldsymbol{I}-\hat{\boldsymbol{V}}\boldsymbol{D}_y)'$, where $\boldsymbol{D}_y = \partial(\log\hat{\boldsymbol{y}})/\partial\boldsymbol{y}'$ is given by*

$$(16) \qquad \boldsymbol{D}_y = \boldsymbol{X}(\boldsymbol{X}'\boldsymbol{W}\hat{\boldsymbol{V}}\boldsymbol{X})^{-1}\boldsymbol{X}'\boldsymbol{W}.$$

The above covariance estimates become more accurate as $N$ increases.

If the test statistics are mildly correlated, the Poisson regression scheme may still be used to fit the empirical null, but the covariance estimates need to change. In this case, the delta-method covariance formulas in Propositions 1 and 2 are applied with $\hat{\boldsymbol{V}}_N$ replaced by a covariance estimate other than (13) that reflects the correlation between the bin counts. This is illustrated below in the DTI example using permutations. Alternatively, one may fit the empirical null including an overdispersion parameter in the Poisson regression. The overdispersion parameter $\phi$ is estimated by the quasi-likelihood MLE $\hat{\phi} = (1/K)\sum_{k=1}^{K}(y_k - \hat{\lambda}_k)/\hat{\lambda}_k$. The Poisson regression fit is the same as before, but the covariance estimates above are inflated by a factor $\phi$.

2.5. *The SNP data.* Recall the SNP data set described in Section 1. The histogram in Figure 1(a) was constructed using bins of width $\Delta = 0.1$ starting from zero. The empirical null was obtained using $\chi^2$ mode matching (Appendix A.2). The fitting interval was defined as $S_0 = [0, 20]$, wide enough to use most of the data without including the far tail region $t > 20$, where discoveries are likely to be made. These choices are discussed in Section 4.

The estimated parameters $\boldsymbol{\theta}^+$ are listed in Table 1. Assuming independence of the test statistics, the associated standard errors (SE) were computed as the square root of the diagonal of (15) using the multinomial covariance (13). For comparison, I also used the bootstrap as follows. Again assuming independence, repeated resampling with replacement from $\{T_1, \ldots, T_N\}$ gave sets $\{T_1^*, \ldots, T_N^*\}$, each leading to a parameter estimate $(\hat{\boldsymbol{\theta}}^+)^*$. The bootstrap covariance estimate of $\hat{\boldsymbol{\theta}}^+$ was computed as the empirical covariance of the $(\hat{\boldsymbol{\theta}}^+)^*$ and the SEs as the square roots of the diagonal elements of this covariance. Notice that the delta-method SEs are only slightly smaller than the bootstrap SEs.

The CIs for $a$ and $\nu$ do not include the theoretical values 1 and 4, indicating a significant departure from the theoretical null in both scaling and degrees of freedom. The CI for $\log(p_0)$ includes 0. This does not prove that there are no significant SNPs, but it shows that the study may not have enough power to discover them. The lower bound of the CI for $\log(p_0)$ suggests that the fraction of non-null SNPs may be as high as $1 - \hat{p}_0 \approx -\log(\hat{p}_0) = 3.18 \times 10^{-5}$, which is about 3 SNPs out of $N = 95{,}810$. If instead of fitting the full empirical null, $p_0$ is estimated alone believing the theoretical null, the result is $\log(\hat{p}_0) = 1.24 \times 10^{-4}$ with standard CI $[1.96 \times 10^{-5}, 2.28 \times 10^{-4}]$. The theoretical null does not admit an estimate of $p_0$ that is less than 1, again indicating that the theoretical null is unsuitable for this data.

The SEs in Table 1 are smaller than they would be if the dependence between the test statistics were taken into account. I did not take into account the dependence because I did not have access to the original data but only to the FBAT test statistics. Given the complexity of the FBAT procedure, pairwise correlations between test statistics would be hard to estimate. It has been claimed that the SNPs in this dataset are not highly correlated because of their widespread locations on the genome [Herbert et al. (2006)]. One indication that the correlation may not have a large effect is that the estimated overdispersion from fitting the empirical null is $\hat{\phi} = 1.090$ (bootstrap SE $= 0.066$), not significantly larger than 1.

2.6. *The DTI data.* For the DTI data set, the histogram in Figure 2(a) was constructed using bins of width $\Delta = 0.05$ starting from zero and the empirical null was obtained using $\chi^2$ mode matching (Appendix A.2). The fitting interval was defined as $S_0 = [0, 4.5]$. These choices are discussed in Section 4.

The estimated parameters $\boldsymbol{\theta}^+$ are listed in Table 2. The associated SEs were computed as the square root of the diagonal of (15) using two different values for $\hat{\boldsymbol{V}}_N$. The naive estimate (column 4) assumes independence of the test statistics and uses the multinomial covariance (13). The estimate in column 5 replaces $\hat{\boldsymbol{V}}_N$ by a permutation estimate $\hat{\boldsymbol{V}}_P$ similar to the one used by Efron (2007a), obtained as follows. In the original data, the group labels of the two groups of 6 subjects were permuted, for a total of $924/2 = 462$ distinct permutations (the test statistics are symmetric, yielding the same value if the groups are swapped). For each permutation, the 20,931 test statistics $T^*$ were recomputed and a vector $\boldsymbol{y}^*$ of histogram counts was produced. The permutation covariance estimate $\hat{\boldsymbol{V}}_P$ was computed as the empirical covariance of the $\boldsymbol{y}^*$.

TABLE 1
*SNP example: Theoretical null parameters (column 2) and empirical null estimates (column 3). Included are delta-method and bootstrap SE (columns 4 and 5) and standard 95% confidence intervals based on the delta-method SE (column 6)*

| $\boldsymbol{\theta}^+$ | Theory | $\hat{\boldsymbol{\theta}}^+$ | SE (15) | Bootstrap SE | 95% CI |
|---|---|---|---|---|---|
| $\log(p_0)$ | 0 | $7.49 \times 10^{-5}$ | $5.45 \times 10^{-5}$ | $5.89 \times 10^{-5}$ | $[-3.18, 18.2] \times 10^{-5}$ |
| $a$ | 1 | 0.9509 | 0.0046 | 0.0048 | $[0.9419, 0.9600]$ |
| $\nu$ | 4 | 4.2675 | 0.0183 | 0.0199 | $[4.2316, 4.3034]$ |

This permutation scheme relies on the subjects being independent but preserves the correlation structure between the test statistics. While validity of the permutations requires the complete null assumption, removal of the mean effects is difficult in this case because of the directional nature of the data. Yet, since $p_0$ is close to 1, the null hypothesis is valid in most voxels and may be enough for the purposes of estimating global parameters, such as those of the empirical null. Table 2 shows that taking into account correlation via the permutation scheme about doubles the naive SEs.

Based on the permutation SEs, the CI for $a$ includes the theoretical value 1, while the CI for $\nu$ does not include the theoretical value 2. This indicates a significant departure from the theoretical null in degrees of freedom but not scaling. The CI for $p_0$ does not include 1 and it is estimated that there are $(1-p_0)N = 745$ non-null voxels in the data. The estimated overdispersion is $\hat{\phi} = 1.332$ (permutation SE = 0.087), significantly larger than 1 as expected from the dependence.

If instead of fitting the full empirical null, $p_0$ is estimated alone believing the theoretical null (blue curve in Figure 2), the result is $\hat{p}_0 = 0.989$ with naive CI $[0.984, 0.994]$ and permutation CI $[0.955, 1.024]$. The theoretical null not only provides a poor fit to the data but gives a higher estimate of $p_0$ than the empirical null, so one may say it is less powerful.

## 3. FDR inference.

TABLE 2

*DTI example: Theoretical null parameters (column 2) and empirical null estimates (column 3). Included are delta-method SEs assuming independence (column 4) and using permutations (column 5). The standard 95% confidence intervals are based on the permutation SEs (column 6)*

| $\theta^+$ | Theory | $\hat{\theta}^+$ | SE (15), ind | SE (15), perm | 95% CI, perm |
|---|---|---|---|---|---|
| $p_0$ | 1 | 0.964 | 0.0042 | 0.0080 | $[0.949, 0.981]$ |
| $a$ | 1 | 0.961 | 0.0225 | 0.0679 | $[0.828, 1.094]$ |
| $\nu$ | 2 | 1.817 | 0.0223 | 0.0377 | $[1.743, 1.891]$ |

3.1. *FDR estimates.* Mode matching is particularly convenient for FDR estimation, as FDR estimates follow immediately from the Poisson regression fits (9). Let $F(t) = p_0 F_0(t) + (1-p_0) F_A(t)$ be the cumulative version of (1). Recall that the local false discovery rate (fdr) and (positive) right-tail FDR are given respectively by

$$(17) \quad \text{fdr}(t) = \frac{p_0 f_0(t)}{f(t)}, \qquad \text{Fdr}_R(t) = \frac{p_0(1 - F_0(t))}{1 - F(t)} = \frac{\int_t^\infty \text{fdr}(u) f(u)\, du}{\int_t^\infty f(u)\, du}$$

[Efron et al. (2001), Efron (2004)]. Using (3) and (4), the local fdr at the bin centers $t_k$ is estimated by

$$(18) \quad \widehat{\text{fdr}}(t_k) = \frac{\hat{p}_0 \hat{f}_0(t_k)}{\hat{f}(t_k)} = \frac{\hat{\lambda}_k/(N\Delta)}{y_k/(N\Delta)} = \frac{\hat{y}_k}{y_k},$$

defined whenever $y_k > 0$, or in vector form as

$$(19) \quad \log \widehat{\mathbf{fdr}} = \log \hat{\boldsymbol{y}} - \log \boldsymbol{y},$$

where $\hat{\boldsymbol{y}}$ is given by (9). In contrast to Efron (2007b), I estimate $\text{Fdr}_R$ at the bin centers $t_k$ based on the right side of (17) by

$$(20) \quad \widehat{\text{Fdr}}_R(t_k) = \frac{(1/2)\widehat{\text{fdr}}(t_k)\hat{f}(t_k) + \sum_{j=k+1}^K \widehat{\text{fdr}}(t_j)\hat{f}(t_j)}{(1/2)\hat{f}(t_k) + \sum_{j=k+1}^K \hat{f}(t_j)}$$

$$= \frac{(1/2)\hat{y}_k + \sum_{j=k+1}^K \hat{y}_j}{(1/2) y_k + \sum_{j=k+1}^K y_j},$$

where (3) and (18) were used. This can be written in vector form as

$$(21) \quad \log \widehat{\mathbf{Fdr}}_R = \log(\boldsymbol{S}\hat{\boldsymbol{y}}) - \log(\boldsymbol{S}\boldsymbol{y}),$$

where $\boldsymbol{S}$ is an upper triangular matrix with entries $1/2$ on the diagonal and 1 above the diagonal. The estimate (21) is easy to analyze theoretically in terms of bias (see Section 4). For the left tail FDR, definition (17) is changed to $\text{Fdr}_L(t) = p_0 F_0(t)/F(t)$ and is estimated similarly by

$$(22) \quad \log \widehat{\mathbf{Fdr}}_L = \log(\boldsymbol{S}'\hat{\boldsymbol{y}}) - \log(\boldsymbol{S}'\boldsymbol{y}),$$

where $\boldsymbol{S}'$ is the transpose of $\boldsymbol{S}$, a lower triangular matrix with entries $1/2$ on the diagonal and 1 below the diagonal.

PROPOSITION 3. (a) *The delta method covariance estimate of the local fdr (19) is* $\widehat{\text{cov}}(\log \widehat{\mathbf{fdr}}) = \boldsymbol{A}\hat{\boldsymbol{V}}_N \boldsymbol{A}'$, *where* $\boldsymbol{A} = \partial(\log \widehat{\mathbf{fdr}})/\partial \boldsymbol{y}' = \boldsymbol{D}_y - \boldsymbol{V}^{-1}$, $\boldsymbol{V} = \text{Diag}(\boldsymbol{y})$ *and* $\boldsymbol{D}_y$ *is given by (16).*

(b) *The delta method covariance estimate of the right tail FDR (21) is* $\widehat{\text{cov}}(\log \widehat{\mathbf{Fdr}}_R) = \boldsymbol{B}\hat{\boldsymbol{V}}_N \boldsymbol{B}'$, *where* $\boldsymbol{B} = \partial(\log \widehat{\mathbf{Fdr}}_R)/\partial \boldsymbol{y}' = \hat{\boldsymbol{U}}^{-1}\boldsymbol{S}\hat{\boldsymbol{V}}\boldsymbol{D}_y - \boldsymbol{U}^{-1}$ *and* $\boldsymbol{U} = \text{Diag}(\boldsymbol{S}\boldsymbol{y})$, $\hat{\boldsymbol{U}} = \text{Diag}(\boldsymbol{S}\hat{\boldsymbol{y}})$. *The formula for the left tail FDR (22) has the same form with $\boldsymbol{S}$ replaced by $\boldsymbol{S}'$.*

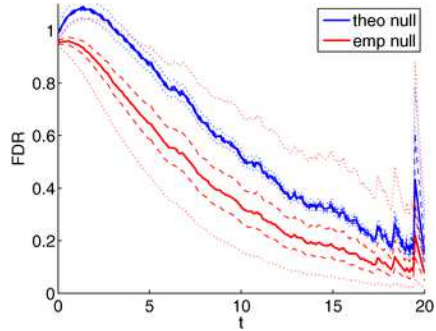

Fig. 3. *DTI example: Right tail FDR estimates using both the theoretical (blue) and the empirical null (red). Included are pointwise standard 95% CIs using the naive multinomial covariance estimate (dashed) and the permutation covariance estimate (dotted).*

As with the empirical null covariance estimates of Section 2.4, the FDR covariance estimates above become more accurate as $N$ increases.

3.2. *The DTI data.* Figure 3 shows the right tail FDR for the DTI data, computed by exponentiating (21). The CIs were computed by exponentiating the CIs for $\log \widehat{\mathbf{Fdr}}_R$, based on SEs equal to the square roots of the diagonal elements of the delta-method covariance estimates given by Proposition 3(b). The narrow CIs correspond to the naive multinomial covariance (13), while the wide CIs correspond to the permutation covariance $\hat{\boldsymbol{V}}_P$ described in Section 2.6. In Schwartzman, Dougherty and Taylor (2008a) it was claimed that the empirical null was more powerful because the FDR curve is always below the FDR curve for the theoretical null, yielding lower thresholds. However, Figure 3 shows that when the dependence is taken into account by using the permutation covariance, the empirical null FDR curve has too much variance. While the estimate of $p_0$ found before indicates that there are non-null voxels in this data, the variance of the FDR estimates makes the results inconclusive.

I defer the FDR analysis of the SNP data to after the following discussion on bias.

**4. Tuning parameters and bias.** Mode matching is controlled by two tuning parameters, the bin width $\Delta$ and the fitting interval $S_0$. Each is connected to a different source of bias: the bin width $\Delta$ controls the bias incurred by using a first order approximation in (2); the fitting interval $S_0$ controls the bias incurred by the inclusion of the alternative component $(1-p_0)f_A$ in the fit of the empirical null. In what follows, I refer to the following two simulation scenarios of model (1):

(23)    $T_i \stackrel{\text{ind}}{\sim} \begin{cases} f_0 = N(0.2, 1.2^2), & \text{probability } p_0, \\ f_A = N(3, 1.2^2), & \text{probability } 1 - p_0, \end{cases}$

(24)     $T_i \overset{\text{ind}}{\sim} \begin{cases} f_0 = 0.8\chi^2(3), & \text{probability } p_0, \\ f_A = \text{noncentral}\chi^2(3, \delta = 3), & \text{probability } 1 - p_0, \end{cases}$

where $\delta = 3$ denotes the noncentrality parameter. The fitting interval is set to $S_0 = [0.2 - t_0, 0.2 + t_0]$ in the normal case and $S_0 = [0, t_0]$ in the $\chi^2$ case, so that in both cases $S_0$ is tuned by the single number $t_0$.

4.1. *The bin width.* The first and smallest source of bias is the use of a first order approximation in (2). Under the zero assumption, the error may be approximated by the next expansion term

(25)     $\pi_k - f_0(t_k)\Delta \approx \frac{f_0''(t_k)}{24}\Delta^3, \qquad t_k \in S_0.$

As in nonparametric density estimation, bias is reduced by thinning the bins. However, the size of the bias depends also on the curvature of the empirical null. If $f_0$ is $N(\mu, \sigma^2)$, the largest curvature occurs at the mode $\mu$, where $f_0''(\mu) \approx -0.4/\sigma^3$. Thus, the error (25) is bounded by $0.017\Delta^3/\sigma^3$. This is about $1.3 \times 10^{-4}$ if $\Delta = 0.2\sigma$. If $f_0$ is $a\chi^2(\nu)$, the curvature depends strongly on $\nu$. For $\nu < 6$ except $\nu = 2$, the curvature is unbounded at $t = 0$, but it decreases rapidly as $t$ increases away from 0. For $\nu \geq 2$, more important is the curvature at the mode, where most of the probability mass lays. This curvature decreases rapidly as $\nu$ increases away from 2. For example, for $\nu = 3$, the curvature at the mode $\nu - 2$ is $f_0''(\nu - 2) \approx -0.12/a^3$ so the error (25) is bounded by $0.005\Delta^3/a^3$. This is about $4 \times 10^{-5}$ if $\Delta = 0.2a$.

The effect of $\Delta$ is illustrated in Figure 4. The plotted empirical null estimates are averages over 100 simulated instances of models (23) and (24) with $p_0 = 1$, $N = 10{,}000$, and fixed $t_0 = 1$ in the normal case and $t_0 = 4$ in the $\chi^2$ case. In contrast to nonparametric density estimation, the variance is remarkably insensitive to $\Delta$. Within the plotted range, the variance is smallest for the smallest value of $\Delta$. This is because, for fixed $S_0$, a small $\Delta$ implies a large number of bins $K = |S_0|/\Delta$ and thus a large number of design points for the Poisson regression. Therefore, one may want the smallest possible $\Delta$ as long as the number of counts $y_k$ within each bin in $S_0$ remains large, say, in the hundreds. Of course, this depends on the number of tests $N$. A larger $N$ allows a smaller $\Delta$. The caveat is computational. A large number of bins $K$ implies inverting large matrices in the fitting of the empirical null. For $N = 10{,}000$, based on Figure 4 and computational considerations, $\Delta = 0.05 \sim 0.1$ seems a reasonable choice for the normal and $\chi^2(\nu)$ with $\nu \geq 2$. If $\nu < 2$, the curvature near $t = 0$ demands much smaller values of $\Delta$ to avoid substantial bias. For large $\nu$, the increase in the effective support of the density may require increasing $\Delta$ in order to reduce computations.

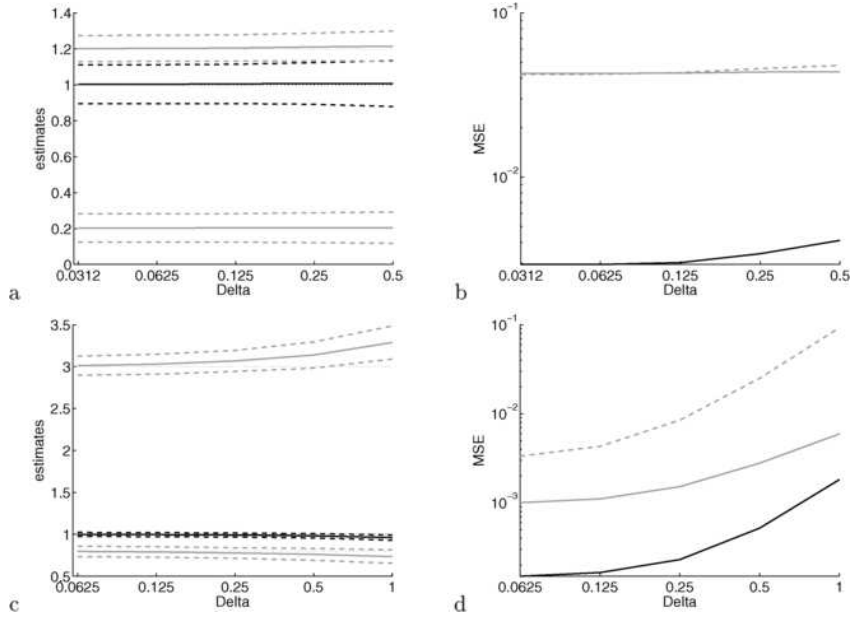

FIG. 4. *Effect of the bin width $\Delta$. Top panels: normal simulation.* (a) *Estimates of $p_0$ (black solid), $\mu$ (lower gray solid) and $\sigma$ (upper gray solid). The thick dashed lines are simulated $\pm 2$ standard errors, very close to the values predicted by formula* (15). *The thin dashed lines indicate the true values $p_0 = 1$, $\mu_0 = 0.2$, $\sigma_0 = 1.2$.* (b) *Simulated MSE for $p_0$ (black), $\mu$ (solid gray) and $\sigma$ (dashed gray). Bottom panels: $\chi^2$ simulation.* (c) *Estimates of $p_0$ (black solid), $a$ (lower gray solid) and $\nu$ (upper gray solid). The thick dashed lines indicate simulated $\pm 2$ standard errors, very close to the values predicted by the formula* (15). *The thin dashed lines indicate the true values $p_0 = 1$, $a_0 = 0.8$, $\nu_0 = 3$.* (d) *Simulated MSE for $p_0$ (black), $a$ (solid gray) and $\nu$ (dashed gray).*

4.2. *The fitting interval.* The largest source of bias in the estimation of the null density is the inclusion of the alternative component $(1 - p_0) f_A$ within $S_0$. The asymptotic bias for large $N$ is quantified in the following proposition.

PROPOSITION 4. *Define the vectors $\boldsymbol{f}_0 = f_0(\boldsymbol{t})$, $\boldsymbol{f}_A = f_A(\boldsymbol{t})$, and let $\hat{\boldsymbol{\eta}}_\infty^+$ and $\hat{\boldsymbol{\theta}}_\infty^+$ denote the deterministic limits of the estimators $\hat{\boldsymbol{\eta}}^+$ and $\hat{\boldsymbol{\theta}}^+$ as $N \to \infty$. The asymptotic biases of $\hat{\boldsymbol{\eta}}_\infty^+$ and $\hat{\boldsymbol{\theta}}_\infty^+$ are given approximately by*

$$
\begin{aligned}
(26) \quad \hat{\boldsymbol{\eta}}_\infty^+ - \boldsymbol{\eta}^+ &\approx (1 - p_0)(\boldsymbol{X}'\boldsymbol{W}\operatorname{Diag}(\boldsymbol{f}_0)\boldsymbol{X})^{-1}\boldsymbol{X}'\boldsymbol{W}(\boldsymbol{f}_A - \boldsymbol{f}_0) \\
&\quad - (\log(p_0), \boldsymbol{0}')'
\end{aligned}
$$

*and $\hat{\boldsymbol{\theta}}_\infty^+ - \boldsymbol{\theta}^+ \approx \boldsymbol{D}(\hat{\boldsymbol{\eta}}_\infty^+ - \boldsymbol{\eta}^+)$, where $\boldsymbol{D}$ is given by* (15).

Roughly, the asymptotic bias (26) is proportional to the likelihood ratio between $f_A$ and $f_0$ within $S_0$, but modified by the Poisson regression. For fixed $\Delta$, the fitting interval controls the number of columns of $\boldsymbol{X}$. Obviously the bias is zero if $p_0 = 1$. The approximation (26) is valid for $p_0$ close to 1. Its accuracy is shown in Table 3 for $p_0 = 0.9$.

TABLE 3
Asymptotic bias in the estimation of $\boldsymbol{\theta}^+$. The limit $\hat{\boldsymbol{\theta}}_\infty^+ = \boldsymbol{\theta}(\hat{\boldsymbol{\eta}}_\infty^+)$ was computed from the solution to the limiting score equation (38)

| Null | $\boldsymbol{\theta}^+$ | $\hat{\boldsymbol{\theta}}_\infty^+$ | $\hat{\boldsymbol{\theta}}_\infty^+ - \boldsymbol{\theta}^+$ | Formula (26) |
|---|---|---|---|---|
| Normal | $\log(p_0) = -0.1054$ | $-0.0801$ | $0.0253$ | $0.0289$ |
|  | $\mu = 0.2$ | $0.2265$ | $0.0265$ | $0.0250$ |
|  | $\sigma^2 = 1.44$ | $1.4907$ | $0.0507$ | $0.0480$ |
| $\chi^2$ | $\log(p_0) = -0.1054$ | $-0.0331$ | $0.0723$ | $0.0749$ |
|  | $a = 0.8$ | $0.8449$ | $0.0449$ | $0.0457$ |
|  | $\nu = 3$ | $2.9789$ | $-0.0211$ | $-0.0210$ |

Figure 5 shows the empirical null parameter estimates averaged over 100 instances of the simulations (23) and (24) with $p_0 = 0.9$ and $N = 10{,}000$. The simulations were repeated for varying $t_0$ and fixed $\Delta = 0.1$. Increasing $t_0$ increases the bias due to the inclusion of the alternative component. On the other hand, increasing $t_0$ also increases the number of design points for the Poisson regression, reducing variance. The bias is worse in the $\chi^2$ simulation because the null and the alternative densities overlap more than in the normal simulation, even though they have a similar separation in their mean. All parameters except the d.f. $\nu$ of the $\chi^2$ tend to be biased upward. This implies that the empirical null is conservative, predicting a smaller contribution of the alternative density in the mixture than there is.

The choice of $t_0$ is more difficult than that of $\Delta$. Ideally, one may want the largest $t_0$ that does not result in a substantial bias. However, the bias depends on $p_0$ and on the alternative density $f_A(t)$, both of which are unknown. The MSE plots in Figure 5 show that in the normal simulation, the optimal $t_0$ is in the range $1.4 \sim 1.9$, corresponding to about $1.2 \sim 1.6$ standard deviations of the true null $N(0.2, 1.2^2)$. Notice that the optimal $t_0$ is not the same for all the parameters. In the $\chi^2$ simulation, the optimal $t_0$ is in the range $3.2 \sim 6$, corresponding to the $74 \sim 94$ percentiles of the true null $0.8\chi^2(3)$. In the DTI example (Section 2.6), I chose $t_0 = 4.5$, which corresponds roughly to the 89.5 percentile of the $\chi^2(2)$ distribution. In the SNP example (Section 2.5), $p_0$ is extremely close to 1, allowing $t_0$ to be much larger. There I used $t_0 = 20$, which is the 99.95 percentile of the $\chi^2(4)$ distribution.

4.3. *Bias in FDR estimation.* One issue overlooked by Efron (2007b) is that the local fdr estimate (18) is biased by definition. The bias is given by the following proposition.

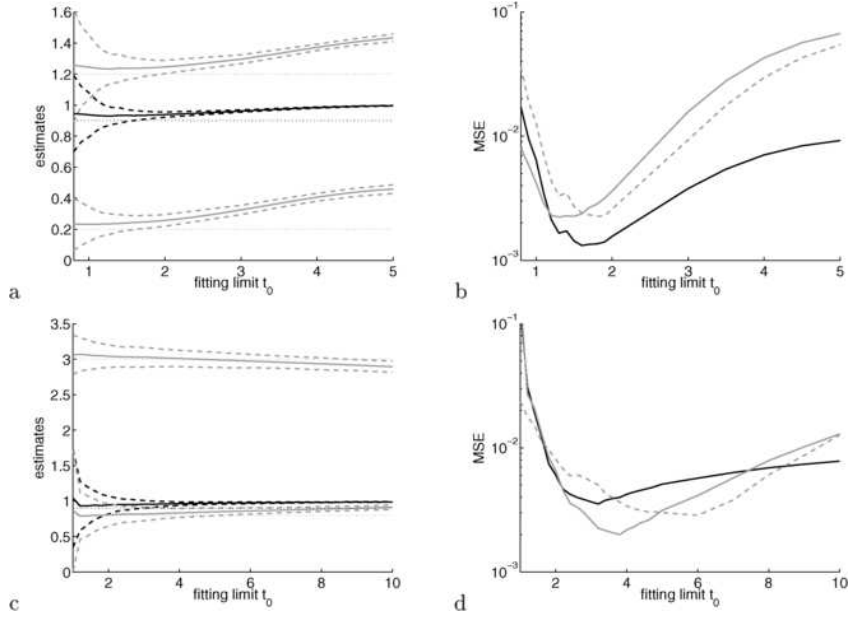

FIG. 5. *Effect of the fitting interval width $t_0$. Top panels: Normal simulation.* (a) *Estimates of $p_0$ (black solid), $\mu$ (lower gray solid) and $\sigma$ (upper gray solid). The thick dashed lines indicate simulated $\pm 2$ standard errors, very close to the values predicted by the formula (15). The thin dashed lines indicate the true values $p_0 = 0.9$, $\mu_0 = 0.2$, $\sigma_0 = 1.2$.* (b) *Simulated MSE for $p_0$ (black), $\mu$ (solid gray) and $\sigma$ (dashed gray). Bottom panels: $\chi^2$ simulation.* (c) *Estimates of $p_0$ (black solid), a (lower gray solid) and $\nu$ (upper gray solid). The thick dashed lines indicate simulated $\pm 2$ standard errors, very close to the values predicted by the formula (15). The thin dashed lines indicate the true values $p_0 = 0.9$, $a_0 = 0.8$, $\nu_0 = 3$. Notice the vertical scale is different from panel* (a). (d) *Simulated MSE for $p_0$ (black), a (solid gray) and $\nu$ (dashed gray).*

PROPOSITION 5. (a) *If $p_0$ and $f_0$ are known, then under the Poisson model (4), $\mathrm{E}[\widehat{\mathrm{fdr}}_k | y_k > 0] = \mathrm{fdr}_k \ \zeta(\lambda_k)$, where*

$$\zeta(\lambda) = \frac{\lambda}{e^\lambda - 1} \int_0^\lambda \frac{e^u - 1}{u} \, du. \tag{27}$$

(b) *If $p_0$ and $f_0$ are estimated by mode matching, then, for large $N$ and $t_k \notin S_0$, $\mathrm{E}[\widehat{\mathrm{fdr}}_k | y_k > 0] \approx \mathrm{fdr}_k \, b_k \zeta(\lambda_k)$, where $b_k$ is the kth entry of the asymptotic bias vector $\boldsymbol{b}_\infty = \exp(\boldsymbol{X}(\hat{\boldsymbol{\eta}}_\infty^+ - \boldsymbol{\eta}^+))$ with the inner parenthesis given by (26).*

The bias factor (27) appears because for the local fdr denominator $y_k \sim Po(\lambda_k)$, $1/y_k$ is biased for $1/\lambda_k$. Figure 6(a) shows a plot of the function $\zeta(\lambda)$, closely related to the so-called exponential integral [Abramowitz and Stegun (1966), Chapter 5]. Since $\lambda_k$ increases with $\gamma$ (i.e., $N$), most bins fall on the right end of the plot, making the local fdr estimate in most bins slightly conservatively biased above the correct value. However, $\lambda_k$ is always small in the far tails of the null density $f_0$, making the FDR estimate biased down. This can be misleading when the true FDR is close to 1.

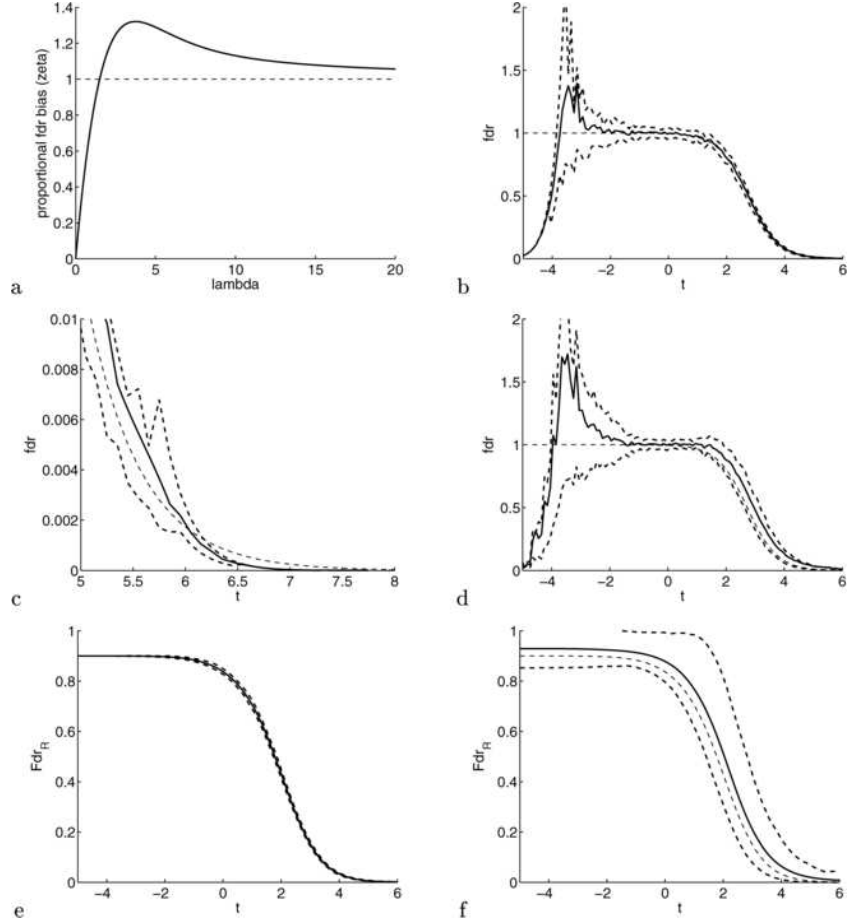

FIG. 6. *FDR bias in the normal simulation.* (a) *The proportional FDR bias function* $\zeta(\lambda)$. *In plots* (b)–(f), *the black solid line is the simulated average estimate, the thick dashed lines are 5 and 95 percentiles of the simulation, and the thin dashed line is the true value.* (b) *Local fdr using theoretical null.* (c) *Local fdr using theoretical null (zoom in).* (d) *Local fdr using empirical null.* (e) *Right tail FDR using theoretical null.* (f) *Right tail FDR using empirical null.*

Figure 6(b) shows the local fdr estimates for 100 instances of the normal simulation (23) with known $p_0 = 0.9$ and $N = 10{,}000$. Because the alternative density sits on the right side of the plot, the true local fdr at the left tail is 1. The local fdr estimate, however, follows the graph of $\zeta(\lambda)$, as the bin counts get smaller toward the left, with the variance proportional to that graph. Any particular realization of the FDR curve here may give the impression that there is something to discover at the left tail, while in reality there is not. The zoom-in in panel c shows that the phenomenon still occurs in the right tail, as the average FDR estimate is first biased up with higher variance and then dips below the truth as the bin counts get very small. This is not noticeable in panel b because, by Proposition 5, the bias is proportional to the true FDR, which is low in this region. When the empirical null is used, the additional bias and variance are visible in panel d. The additional bias is captured by the factor $b_k$ of Proposition 5(b), where the approximation is the result of using the asymptotic bias for large $N$ derived from Proposition 4. Notice in Figure 6(d) that the bias is up, so the FDR estimates are conservative.

The tail FDR also suffers from a similar bias phenomenon, being sensitive to small cumulative bin counts in the far tails. Following a similar argument as in Proposition 5(a), when $p_0$ and $f_0$ are known, (20) says $\widehat{\text{Fdr}}_k = (\boldsymbol{S}\boldsymbol{\lambda}_0)_k/(\boldsymbol{S}\boldsymbol{y})_k$, where the cumulative denominator $y_k/2 + \sum_{l=k+1}^{K} y_l$ behaves similarly to a Poisson random variable with mean $(\boldsymbol{S}\boldsymbol{\lambda})_k = \lambda_k/2 + \sum_{l=k+1}^{K} \lambda_l$. Therefore, $\text{E}[\widehat{\text{Fdr}}_k|y_k > 0] = \text{Fdr}_k \cdot \text{E}[(\boldsymbol{S}\boldsymbol{\lambda})_k/(\boldsymbol{S}\boldsymbol{y})_k|(\boldsymbol{S}\boldsymbol{y})_k > 0]$, where the conditional expectation behaves approximately like $\zeta[(\boldsymbol{S}\boldsymbol{\lambda})_k]$. A simulation using $p_0 = 1$ (not shown) gives FDR curves that are very similar to those on the left end of panels b and d. In Figure 6(e) (zoom-in not shown), the bias is visible but small because the FDR itself is low in the right tail. Panel f shows the increase in bias and variance when the empirical null is used.

The bias phenomenon does not contradict the results of Storey, Taylor and Siegmund (2004), which claim asymptotic unbiasedness of the tail FDR estimator. Consider a fixed bin $k$. As $N$ increases, the expected bin count $\lambda_k$ increases and the operating point in Figure 6(a) moves to the right, making the FDR estimate asymptotically unbiased. The $\zeta(\lambda)$ phenomenon appeals to practical cases where $N$ is large but finite, so that the bin counts at the tails are still small.

Efron (2004, 2007b) circumvented the bias problem by smoothing the histogram with a spline fit. This helps because the compounding of data at each bin pushes the operating point in the $\zeta(\lambda)$ graph [Figure 6(a)] to the right. However, smoothing introduces a bias of its own. Simulations using smoothing show that the resulting estimates at the far tails, especially when $p_0 = 1$, are not reliable, as they are very sensitive to the choice of knots or smoothing bandwidth.

4.4. *The SNP data.* The FDR analysis is summarized in Figure 7. Here I focus on the local fdr, which in this case is more powerful than the tail FDR. The observed local fdr estimates (18) are compared to their expected value under the complete null, computed by replacing $\lambda_k$ for $\hat{\lambda}_k$ in Proposition 5 and setting $\text{fdr}_k = 1$ and $b_k = 1$. In agreement with Proposition 5, the expected local fdr estimate goes down at the far tails where the number of counts is small. Three bins stand out with observed local fdr estimates significantly below the dashed line. The results for these bins are reported in Table 4. While the observed local fdr values are relatively low, they turn out to be not as low when their bias is taken into account. The adjusted local fdr values in Table 4 are not precise at correcting the bias, but they hint that about 50% of the 8 SNPs contained in these 3 bins might be associated with obesity. This result agrees with the assessment of Section 2.5 that the nonnull distribution may contain about 3 SNPs.

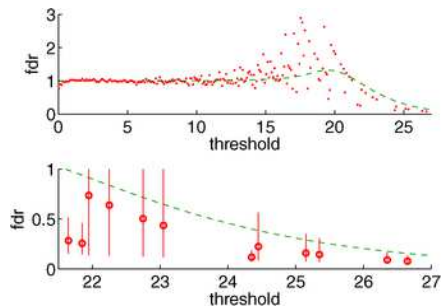

Fig. 7. *SNP example: local fdr estimates using the empirical null. Top panel: full range. Bottom panel: zoom-in including standard 95% CIs. In both panels the dashed line is the expectation of the local fdr estimate under the complete null.*

**5. The effect of correlation.** When the theoretical null is $N(0,1)$, Efron (2007a) showed that the overdispersion in the covariance of the empirical null with respect to the multinomial covariance can be described as a multiple of a "wing function," where the multiplying factor captures the variance of the pairwise correlations between the test statistics. This result extends to exponential families as follows.

TABLE 4
*SNP example: Bins with local fdr significantly below the expected local fdr under the complete null. Column 3 contains standard 95% CIs based on the delta-method SE. Column 4 is the observed local fdr from column 2 divided by the expected local fdr for the complete null at that bin. Column 5 is the number of SNPs in the bin*

| $t_k$ | $\widehat{\text{fdr}}_k$ | 95% CI | Adjusted $\widehat{\text{fdr}}_k$ | # SNPs |
|-------|--------------------------|--------|-----------------------------------|--------|
| 21.65 | 0.2831 | [0.1546, 0.5184] | 0.4169 | 3 |
| 21.85 | 0.2575 | [0.1445, 0.4587] | 0.4082 | 3 |
| 24.35 | 0.1173 | [0.0726, 0.1896] | 0.5310 | 2 |

Suppose the theoretical null $f_0$ is one of the Lancaster distributions, that is, the exponential families normal, gamma, Poisson or negative binomial [Koudou (1998)]. Given two random variables $T_i$ and $T_j$ with correlation $\rho$, these distributions admit a bivariate model where both $T_i$ and $T_j$ have the same marginal density $f_0$ and their joint density is given by

$$(28) \qquad f_0(t_i, t_j; \rho) = f_0(t_i) f_0(t_i) \sum_{n=0}^{\infty} \frac{\rho^n}{n!} L_n(t_i) L_n(t_j),$$

where $L_n(t)$ are the Lancaster orthogonal polynomials with respect to $f_0$: Hermite if $f_0$ is normal, generalized Laguerre if $f_0$ is gamma, Charlier if $f_0$ is Poisson, and normalized Meixner if $f_0$ is negative binomial. In particular, when $f_0$ is normal, the expansion (28) is known as Mehler's formula [Patel and Read (1996), Kotz, Balakrishnan and Johnson (2000)] and is equal to the standard bivariate normal with correlation coefficient $\rho$.

THEOREM 1. *Let $f_0$ be one of the Lancaster distributions and assume that under the complete null every pair of test statistics $(T_i, T_j)$ has a bivariate density given by (28) with marginals $f_0(t)$ and $\mathrm{corr}(T_i, T_j) = \rho_{ij}$. Let $E(\rho^n)$, $n = 1, 2, \ldots$, denote the empirical moments of the $N(N-1)$ correlations $\rho_{ij}$, $i < j$. Then the covariance of the vector of bin counts $\boldsymbol{y}$ is*

$$(29) \qquad \mathrm{cov}(\boldsymbol{y}) = \left[\mathrm{Diag}(\boldsymbol{\lambda}) - \frac{\boldsymbol{\lambda}\boldsymbol{\lambda}'}{N}\right] + \left(1 - \frac{1}{N}\right) \mathrm{Diag}(\boldsymbol{\lambda}) \boldsymbol{\delta} \, \mathrm{Diag}(\boldsymbol{\lambda}),$$

*where*

$$(30) \qquad \boldsymbol{\delta} = \sum_{n=1}^{\infty} \frac{E(\rho^n)}{n!} L_n(\boldsymbol{t}) L_n(\boldsymbol{t})'$$

*and $L_n(\boldsymbol{t})$ denote the Lancaster polynomials evaluated at the vector $\boldsymbol{t}$.*

When $f_0(t)$ is $N(\mu, \sigma^2)$ then (30) becomes

$$\boldsymbol{\delta} = \sum_{n=1}^{\infty} \frac{E(\rho^n)}{n!} H_n\!\left(\frac{\boldsymbol{t} - \mu \boldsymbol{1}}{\sigma}\right) H_n\!\left(\frac{\boldsymbol{t} - \mu \boldsymbol{1}}{\sigma}\right)',$$

where $H_n(\boldsymbol{t})$ are the Hermite polynomials: $H_0(t) = 1$, $H_1(t) = t$, $H_2(t) = t^2 - 1$, and so on. In particular, setting $\mu = 0$, $\sigma = 1$, $E(\rho) = 0$, and truncating the series at $n = 2$ gives precisely the result of Efron (2007a), Theorem 1:

$$(31) \qquad \mathrm{cov}(\boldsymbol{y}) = \left[\mathrm{Diag}(\boldsymbol{\lambda}) - \frac{\boldsymbol{\lambda}\boldsymbol{\lambda}'}{N}\right] + \left(1 - \frac{1}{N}\right) E(\rho^2) \boldsymbol{w}_2 \boldsymbol{w}_2',$$

where $\boldsymbol{w}_2 = \mathrm{Diag}(\boldsymbol{\lambda}) H_2(\boldsymbol{t})/\sqrt{2}$ is the "wing function" vector with components $w_{2,k} = N \Delta f_0(t_k)(t_k^2 - 1)/\sqrt{2}$. The above extension to other Hermite polynomial orders is recognized in Remark E of Efron (2007a) but not precisely formulated.

When $f_0(t)$ is the $a\chi^2(\nu)$ density (11) then (30) becomes

$$\boldsymbol{\delta} = \sum_{n=1}^{\infty} \frac{E(\rho^n)}{n!} \frac{\Gamma(\nu/2)}{\Gamma(\nu/2+n)} L_n^{(\nu/2-1)}\left(\frac{\boldsymbol{t}}{2a}\right) L_n^{(\nu/2-1)}\left(\frac{\boldsymbol{t}}{2a}\right)',$$

where $L_n^{(\nu/2-1)}(t)$ are the generalized Laguerre polynomials of degree $\nu/2-1$:

$$L_0^{(\nu/2-1)}(t) = 1$$
$$L_1^{(\nu/2-1)}(t) = -t + \nu/2$$
$$L_2^{(\nu/2-1)}(t) = t^2 - 2(\nu/2+1)t + (\nu/2)(\nu/2+1)$$

and so on. Here there is no reason to assume $E(\rho) = 0$. A similar approximation to (31) to the first order is

$$(32) \qquad \operatorname{cov}(\boldsymbol{y}) = \left[\operatorname{Diag}(\boldsymbol{\lambda}) - \frac{\boldsymbol{\lambda}\boldsymbol{\lambda}'}{N}\right] + \left(1 - \frac{1}{N}\right) E(\rho) \boldsymbol{w}_1 \boldsymbol{w}_1',$$

where

$$(33) \qquad \boldsymbol{w}_1 = \operatorname{Diag}(\boldsymbol{\lambda}) \sqrt{\frac{\Gamma(\nu/2)}{\Gamma(\nu/2+1)}} L_1^{(\nu/2-1)}\left(\frac{\boldsymbol{t}}{2a}\right)$$

is the corresponding first order "wing function" vector with components $w_{1,k} = N\Delta\sqrt{\Gamma(\nu/2)/\Gamma(\nu/2+1)} f_0(t_k)[t_k/(2a) - \nu/2]$.

5.1. *The DTI data.* Recall the permutation estimate $\hat{\boldsymbol{V}}_P$ of the covariance matrix $\operatorname{cov}(\boldsymbol{y})$ obtained in Section 2.6. For this dataset, the largest eigenvalue of $\boldsymbol{V}_N = \operatorname{Diag}(\boldsymbol{\lambda}) - \boldsymbol{\lambda}\boldsymbol{\lambda}'/N$ is only 4.1% of the largest eigenvalue $\hat{d}_1$ of $\hat{\boldsymbol{V}}_P$, while the second eigenvalue $\hat{d}_2$ of $\hat{\boldsymbol{V}}_P$ is only 6.6% of $\hat{d}_1$. Thus, using (32), we can approximate

$$\hat{\boldsymbol{V}}_P \approx E(\rho)\hat{\boldsymbol{w}}_1\hat{\boldsymbol{w}}_1', \tag{34}$$

where $\hat{\boldsymbol{w}}_1$ is the first-order "wing function" (33) evaluated using the empirical null estimates $\hat{\boldsymbol{\lambda}}$, $\hat{\nu}$ and $\hat{a}$ from Section 2.5. Figure 8 shows the first eigenvector of $\hat{\boldsymbol{V}}_P$ superimposed with $\hat{\boldsymbol{w}}_1$ normalized to unit norm, that is, $\hat{\boldsymbol{w}}_1/\|\hat{\boldsymbol{w}}_1\|$ with $\|\hat{\boldsymbol{w}}_1\|^2 = \sum_{k=1}^{K}\hat{w}_{1,k}^2$. The similarity between the two is akin to the similarity between the 1st eigenvector of the permutation covariance estimate and the 2nd order wing function in the normal example described in Efron (2007a).

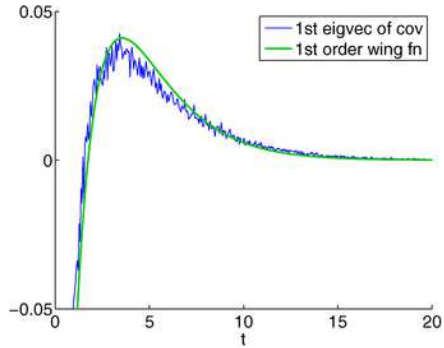

FIG. 8. *DTI example: First eigenvector of the permutation covariance estimate (blue) and first-order $\chi^2$ "wing function" (green).*

Given the equivalence (34), one might be tempted to estimate the average pairwise correlation between the test statistics as $\hat{E}(\rho) = \hat{d}_1\|\hat{\boldsymbol{w}}_1\|^2 = 0.0057$, where $\hat{d}_1$ is the largest eigenvalue of $\hat{\boldsymbol{V}}_P$. This is a strongly conservative estimate, as it suffers from the upward sorting bias of $\hat{d}_1$. The true average pairwise correlation is probably much lower. This is reassuring as the correlation is presumably mostly local in the image domain.

**6. Summary and discussion.** In this article I have extended the central matching method for estimating the null density in large-scale multiple testing to a mode matching method, applicable when the theoretical null belongs to any exponential family or a related distribution such as $t$ or $F$. The empirical null estimate is accompanied by an estimate of $p_0$, the proportion of true null tests in the data. Further, the empirical null estimates can be used directly to estimate local and tail FDR curves for FDR inference. Delta method covariance estimates and bias formulas have been derived. We have seen that FDR estimates are biased down at the far tails and should be taken cautiously whenever the corresponding observed bin counts are small. The effect of correlation has been explained by a generalization of Efron's "wing function."

Efron (2005b, 2007b) discusses several reasons why the empirical null may not match exactly the theoretical null in observational studies. It should be emphasized that mode matching does not necessarily increase power with respect to the theoretical null [Efron (2004) provides counterexamples]. Instead, the empirical null answers a question of model validity.

In the normal case, another empirical null method called MLE fitting [Efron (2007b)] has been reported to give similar empirical null estimates with slightly lower variance. Mode matching is easier to analyze and is appealing because of its application to exponential families. It is also easier to implement in practice because of available software and computational efficiency. But, in principle, just like mode matching, MLE fitting could be extended to other distributions beyond the normal too.

At least two aspects of mode matching may benefit from further study beyond this paper. One aspect is the possibility of choosing data-dependent limits for the fitting interval $S_0$. Fixed limits based on the theoretical null are inappropriate precisely because the empirical null is expected to be displaced or scaled with respect the theoretical null. For instance, in an analysis of $\chi^2$-scores, Schwartzman, Dougherty and Taylor (2008a) used as upper limit the 90th percentile of the empirical test statistic distribution. In the SNP data example above, a bootstrap analysis showed that setting the upper limit to the 99.95th percentile of the test statistic distribution, rather than the 99.95th percentile of the theoretical $\chi^2(4)$ density (the value 20 used previously), results in empirical null variance estimates that are about 50% higher than those obtained when the limit was fixed. This suggests that the cost of a data-dependent limit might not be too high. Unfortunately, as shown in Section 4 above, the choice of the limit depends very much on the alternative distribution and the null proportion, both of which are unknown.

Another aspect is the possibility of using the two-step approach of Efron (2007b) of estimating the mixture density nonparametrically before fitting the empirical null by mode matching. As noted above, the most crucial issue is the bias in the tails of the density. Future exploration may yield an answer to what is the best way to estimate the mixture density for mode matching. For example, different bin widths $\Delta$ could be used inside and outside the fitting interval $S_0$ since they serve different purposes. Inside $S_0$ one could optimize $\Delta$ for empirical null estimation, while outside $S_0$ one could optimize $\Delta$ for FDR estimation.

Matlab functions implementing the methods described in this paper are available at http://biowww.dfci.harvard.edu/~armin/software.html.

## APPENDIX A: SPECIAL CASES

**A.1. Normal family.** The empirical null $N(\mu, \sigma^2)$ with $\boldsymbol{\theta} = (\mu, \sigma^2)$ has exponential family form

$$x(t) = (t, t^2)', \qquad \psi(\boldsymbol{\eta}) = -\frac{\eta_1^2}{4\eta_2} - \frac{1}{2}\log(-2\eta_2)$$

$$\boldsymbol{\eta} = (\eta_1, \eta_2)' = \left(\frac{\mu}{\sigma^2}, -\frac{1}{2\sigma^2}\right)', \qquad g_0(t) = \frac{1}{\sqrt{2\pi}}.$$

The Poisson regression (8) using $t$ and $t^2$ as predictors and $\log(N\Delta/\sqrt{2\pi})$ as offset gives estimates $\hat{\eta}_1$, $\hat{\eta}_2$ and $\hat{C}$. From these we obtain

$$\hat{\mu} = -\frac{\hat{\eta}_1}{2\hat{\eta}_2}, \qquad \hat{\sigma}^2 = -\frac{1}{2\hat{\eta}_2}, \qquad \log \hat{p}_0 = \hat{C} + \psi(\hat{\boldsymbol{\eta}}).$$

The parameter derivative matrix $\hat{\boldsymbol{D}}$ required for computing the covariance (15) of $\hat{\boldsymbol{\theta}}^+ = (\log \hat{p}_0, \hat{\mu}, \hat{\sigma}^2)'$ is

$$\hat{\boldsymbol{D}} = \frac{\partial \hat{\boldsymbol{\theta}}^+}{\partial (\hat{\boldsymbol{\eta}}^+)'} = \begin{pmatrix} 1 & -\frac{\hat{\eta}_1}{2\hat{\eta}_2} & \frac{\hat{\eta}_1^2}{4\hat{\eta}_2^2} - \frac{1}{2\hat{\eta}_2} \\ 0 & -\frac{1}{2\hat{\eta}_2} & \frac{\hat{\eta}_1}{2\hat{\eta}_2^2} \\ 0 & 0 & \frac{1}{2\hat{\eta}_2^2} \end{pmatrix} = \begin{pmatrix} 1 & \hat{\mu} & \hat{\mu}^2 + \hat{\sigma}^2 \\ 0 & \hat{\sigma}^2 & 2\hat{\mu}\hat{\sigma}^2 \\ 0 & 0 & 2\hat{\sigma}^4 \end{pmatrix}.$$

The normal family $N(\mu, \sigma^2)$ lends itself to two exponential subfamilies.

A.1.1. *Estimate* $\theta = \mu$. The empirical null $N(\mu, \sigma_0^2)$ with fixed $\sigma_0^2$ has the exponential family form $x(t) = t$, $\eta = \mu/\sigma_0^2$, $\psi(\eta) = \sigma_0^2 \eta^2/2$ and $g_0(t) = e^{-t^2/(2\sigma_0^2)}/\sqrt{2\pi\sigma_0^2}$. Poisson regression using $t$ as predictor and $\log(N\Delta g_0(t))$ as offset gives estimates $\hat{\eta}$ and $\hat{C}$. From these we obtain

$$\hat{\mu} = \sigma_0^2 \hat{\eta}, \qquad \log \hat{p}_0 = \hat{C} + \frac{\sigma_0^2 \hat{\eta}^2}{2}, \qquad \hat{\boldsymbol{D}} = \begin{pmatrix} 1 & \hat{\mu} \\ 0 & \sigma_0^2 \end{pmatrix}.$$

A.1.2. *Estimate* $\theta = \sigma^2$. The empirical null $N(\mu_0, \sigma^2)$ with fixed $\mu_0$ has the exponential family form $x(t) = (t - \mu_0)^2$, $\eta = -1/(2\sigma^2)$, $\psi(\eta) = (-1/2) \times \log(-2\eta)$ and $g_0(t) = 1/\sqrt{2\pi}$. Poisson regression using $(t - \mu_0)^2$ as predictor and $\log(N\Delta/\sqrt{2\pi})$ as offset gives estimates $\hat{\eta}$ and $\hat{C}$. From these we obtain

$$\hat{\sigma}^2 = -\frac{1}{2\hat{\eta}}, \qquad \log \hat{p}_0 = \hat{C} - \tfrac{1}{2}\log(-2\hat{\eta}), \qquad \hat{\boldsymbol{D}} = \begin{pmatrix} 1 & \hat{\sigma}^2 \\ 0 & 2\hat{\sigma}^4 \end{pmatrix}.$$

**A.2. Scaled $\chi^2$ family (Gamma).** The empirical null $a\chi^2(\nu)$ (11) with $\boldsymbol{\theta} = (a, \nu)'$ has exponential family form

$$x(t) = (t, \log t)', \qquad \psi(\boldsymbol{\eta}) = \log\left(\frac{\Gamma(\eta_2 + 1)}{(-\eta_1)^{\eta_2+1}}\right)$$

$$\boldsymbol{\eta} = (\eta_1, \eta_2)' = \left(-\frac{1}{2a}, \frac{\nu}{2} - 1\right)', \qquad g_0(t) = 1.$$

The Poisson regression (8) using $t$ and $\log t$ as predictors gives estimates $\hat{\eta}_1$, $\hat{\eta}_2$ and $\hat{C}$. From these we obtain

$$\hat{a} = -\frac{1}{2\hat{\eta}_1}, \qquad \hat{\nu} = 2(\hat{\eta}_2 + 1), \qquad \log \hat{p}_0 = \hat{C} + \psi(\hat{\boldsymbol{\eta}}).$$

The parameter derivative matrix $\hat{\boldsymbol{D}}$ required for computing the covariance (15) of $\hat{\boldsymbol{\theta}}^+ = (\log \hat{p}_0, \hat{a}, \hat{\nu})'$ is

$$\hat{\boldsymbol{D}} = \begin{pmatrix} 1 & -\dfrac{\hat{\eta}_2 + 1}{\hat{\eta}_1} & \Psi(\hat{\eta}_2 + 1) - \log(-\hat{\eta}_1) \\ 0 & \dfrac{1}{2\hat{\eta}_1^2} & 0 \\ 0 & 0 & 2 \end{pmatrix}$$

$$= \begin{pmatrix} 1 & \hat{a}\hat{\nu} & \Psi(\hat{\nu}/2) + \log(2\hat{a}) \\ 0 & 2\hat{a}^2 & 0 \\ 0 & 0 & 2 \end{pmatrix},$$

where $\Psi(z) = (d/dz) \log \Gamma(z)$ is the Digamma function. The scaled $\chi^2$ family lends itself to two exponential subfamilies.

A.2.1. *Estimate* $\theta = a$. The empirical null $a\chi^2(\nu_0)$ with fixed $\nu_0$ has exponential family form $x(t) = t$, $\eta = -1/(2a)$, $\psi(\eta) = -(\nu_0/2) \log(-\eta)$ and $g_0(t) = t^{\nu_0/2 - 1}/\Gamma(\nu_0/2)$. Poisson regression using $t$ as a predictor and $\log(N\Delta g_0(t))$ as offset gives estimates $\hat{\eta}$ and $\hat{C}$. From these we obtain

$$\hat{a} = -\frac{1}{2\hat{\eta}}, \qquad \log \hat{p}_0 = \hat{C} - \frac{\nu_0}{2} \log(-\hat{\eta}), \qquad \hat{\boldsymbol{D}} = \begin{pmatrix} 1 & \hat{a}\nu_0 \\ 0 & 2\hat{a}^2 \end{pmatrix}.$$

A.2.2. *Estimate* $\theta = \nu$. The empirical null $a_0\chi^2(\nu)$ with fixed $a_0$ has exponential family form $x(t) = \log t$, $\eta = \nu/2 - 1$, $\psi(\eta) = \log \Gamma(\eta + 1) + (\eta + 1) \log(2a_0)$ and $g_0(t) = e^{-t/(2a_0)}$. Poisson regression using $\log t$ as a predictor and $\log(N\Delta g_0(t))$ as offset gives estimates $\hat{\eta}$ and $\hat{C}$. From these we obtain

$$\hat{\nu} = 2(\hat{\eta} + 1), \qquad \log \hat{p}_0 = \hat{C} + \psi(\hat{\eta}), \qquad \hat{\boldsymbol{D}} = \begin{pmatrix} 1 & \Psi(\hat{\nu}/2) + \log(2a_0) \\ 0 & 2 \end{pmatrix}.$$

## APPENDIX B: PROOFS

PROOF OF PROPOSITION 1. The score equation for the Poisson regression (8) including the external weights $\boldsymbol{W}$ is

(35) $$\boldsymbol{X}'\boldsymbol{W}[\boldsymbol{y} - \exp(\boldsymbol{X}\hat{\boldsymbol{\eta}}^+ + \boldsymbol{h})] = 0.$$

The rate of change of the MLE vector $\hat{\boldsymbol{\eta}}^+$ with respect to the count vector $\boldsymbol{y}$, considered as continuous, is

(36) $$\frac{\partial \hat{\boldsymbol{\eta}}^+}{\partial \boldsymbol{y}'} = (\boldsymbol{X}'\boldsymbol{W}\hat{\boldsymbol{V}}\boldsymbol{X})^{-1}\boldsymbol{X}'\boldsymbol{W},$$

obtained by differentiating (35) with respect to $\boldsymbol{y}$ and replacing (9). Conditional on $N$, the covariance estimate of $\boldsymbol{y}$ is $\hat{\boldsymbol{V}}_N$. Thus, the delta method covariance estimate of $\hat{\boldsymbol{\eta}}^+$ is $(\partial \hat{\boldsymbol{\eta}}^+/\partial \boldsymbol{y}')\hat{\boldsymbol{V}}_N(\partial \hat{\boldsymbol{\eta}}^+/\partial \boldsymbol{y}')'$, yielding (14).

The rate of change of $\boldsymbol{\theta}^+$ with respect to $\boldsymbol{\eta}^+$ is

$$(37) \qquad \boldsymbol{D} = \frac{\partial \boldsymbol{\theta}^+}{\partial (\boldsymbol{\eta}^+)'} = \frac{\partial (\log p_0, \boldsymbol{\theta}(\boldsymbol{\eta})')'}{\partial (C, \boldsymbol{\eta}')} = \begin{pmatrix} 1 & \dot{\psi}(\boldsymbol{\eta})' \\ 0 & \dot{\boldsymbol{\theta}}(\boldsymbol{\eta})' \end{pmatrix},$$

so the rate of change of $\hat{\boldsymbol{\theta}}^+$ with respect to $\hat{\boldsymbol{\eta}}^+$ at $\hat{\boldsymbol{\eta}}$ is $\hat{\boldsymbol{D}} = \boldsymbol{D}(\hat{\boldsymbol{\eta}})$. The delta method covariance estimate of $\hat{\boldsymbol{\theta}}^+$ is $(\partial \hat{\boldsymbol{\theta}}^+/\partial (\hat{\boldsymbol{\eta}}^+)') \widehat{\mathrm{cov}}(\hat{\boldsymbol{\eta}}^+)(\partial \hat{\boldsymbol{\theta}}^+/\partial (\hat{\boldsymbol{\eta}}^+)')'$, yielding (15). □

PROOF OF PROPOSITION 2. The rate of change (16) of the vector $\log \hat{\boldsymbol{y}} = \boldsymbol{X}\hat{\boldsymbol{\eta}}^+$ with respect to $\boldsymbol{y}$ follows directly by (36). By the chain rule, $\partial \hat{\boldsymbol{y}}/\partial \boldsymbol{y}' = [\partial \hat{\boldsymbol{y}}/\partial (\log \boldsymbol{y})'][\partial (\log \boldsymbol{y})/\partial \boldsymbol{y}'] = \hat{\boldsymbol{V}}\boldsymbol{D}_y$, and similarly, $\partial (\boldsymbol{y} - \hat{\boldsymbol{y}})/\partial \boldsymbol{y}' = \boldsymbol{I} - \hat{\boldsymbol{V}}\boldsymbol{D}_y$. The result follows by the delta method. □

PROOF OF PROPOSITION 3. (a) Follows immediately by the delta method and the definition of $\boldsymbol{A}$.

(b) To evaluate the rate of change of the vector $\log(\boldsymbol{S}\hat{\boldsymbol{y}})$ with respect to $\boldsymbol{y}$, compute

$$\frac{\partial (\log(\boldsymbol{S}\hat{\boldsymbol{y}}))_k}{\partial y_l} = \frac{\partial}{\partial y_l} \log\left(\frac{1}{2}\hat{y}_k + \sum_{j=k+1}^K \hat{y}_j\right) = \frac{\frac{1}{2}\frac{\partial \hat{y}_k}{\partial y_l} + \sum_{j=k+1}^K \frac{\partial \hat{y}_j}{\partial y_l}}{\frac{1}{2}\hat{y}_k + \sum_{j=k+1}^K \hat{y}_j}.$$

Thus,

$$\frac{\partial (\log(\boldsymbol{S}\hat{\boldsymbol{y}}))}{\partial \boldsymbol{y}'} = \hat{\boldsymbol{U}}^{-1}\boldsymbol{S}\frac{\partial \hat{\boldsymbol{y}}}{\partial \boldsymbol{y}'} = \hat{\boldsymbol{U}}^{-1}\boldsymbol{S} \cdot \hat{\boldsymbol{V}}\frac{\partial (\log \hat{\boldsymbol{y}})}{\partial \boldsymbol{y}'} = \hat{\boldsymbol{U}}^{-1}\boldsymbol{S}\hat{\boldsymbol{V}}\boldsymbol{D}_y,$$

where we have used the fact that $\partial (\log \hat{y}_k)/\partial y_l = (1/\hat{y}_k)\partial \hat{y}_k/\partial y_l$. The result now follows by the delta method and the definition of $\boldsymbol{B}$. □

PROOF OF PROPOSITION 4. Dividing the score equation (35) by $\gamma \Delta$ and applying the law of large numbers as $\gamma \to \infty$ gives that $\hat{\boldsymbol{\eta}}^+$ converges to the solution $\hat{\boldsymbol{\eta}}^+_\infty$ of the equation

$$(38) \qquad \boldsymbol{X}'\boldsymbol{W}[p_0 \boldsymbol{f}_0 + (1-p_0)\boldsymbol{f}_A - \mathrm{Diag}(g_0(\boldsymbol{t}))\exp(\boldsymbol{X}\hat{\boldsymbol{\eta}}^+_\infty)] = 0.$$

In particular, if $p_0 = 1$, we have that $\hat{\boldsymbol{\eta}}^+$ is asymptotically unbiased, that is, $\hat{\boldsymbol{\eta}}^+_\infty = (0, \boldsymbol{\eta}')'$. The idea is to find a first order expansion of $\hat{\boldsymbol{\eta}}^+$ near $p_0 = 1$. Differentiating (38) with respect to $p_0$, we obtain that, at $p_0 = 1$, $d\hat{\boldsymbol{\eta}}^+_\infty/dp_0 = (\boldsymbol{X}'\boldsymbol{W}\mathrm{Diag}(\boldsymbol{f}_0)\boldsymbol{X})^{-1}\boldsymbol{X}'\boldsymbol{W}(\boldsymbol{f}_0 - \boldsymbol{f}_A)$. The bias in the estimation of $\boldsymbol{\eta}^+$ is approximately

$$\hat{\boldsymbol{\eta}}^+_\infty - \boldsymbol{\eta}^+ = \hat{\boldsymbol{\eta}}^+_\infty - (0, \boldsymbol{\eta}')' - (\log(p_0), \boldsymbol{0}')'$$
$$\approx \left.\frac{d\hat{\boldsymbol{\eta}}^+_\infty}{dp_0}\right|_{p_0=1}(p_0 - 1) - (\log(p_0), \boldsymbol{0}')',$$

yielding (26). Similarly, the bias in the estimation of $\boldsymbol{\theta}^+$ is a first order expansion of $\boldsymbol{\theta}^+$ with respect to $\boldsymbol{\eta}^+$ near $p_0 = 1$. □

PROOF OF PROPOSITION 5. (a) For known $p_0$ and $f_0$, (18) says $\widehat{\text{fdr}}_k = \lambda_{k,0}/y_k$, where $y_k \sim Po(\lambda_k)$, $\lambda_k = \gamma \Delta f(t_k)$ and $\lambda_{k,0} = \gamma \Delta p_0 f_0(t_k)$. Thus,

$$\text{E}[\widehat{\text{fdr}}_k | y_k > 0] = \text{E}\left(\frac{\lambda_{k,0}}{y_k}\Big| y_k > 0\right) = \frac{\lambda_{k,0}}{\lambda_k}\text{E}\left(\frac{\lambda_k}{y_k}\Big| y_k > 0\right) = \text{fdr}_k \cdot \zeta(\lambda_k),$$

where $\zeta(\lambda)$ is defined for a generic $y \sim Po(\lambda)$ as $\zeta(\lambda) = \text{E}(\lambda/y | y > 0)$. By direct evaluation,

$$\zeta(\lambda) = \frac{1}{1 - e^{-\lambda}} \sum_{j=1}^{\infty} \frac{\lambda}{j} \frac{e^{-\lambda}\lambda^j}{j!} = \frac{\lambda}{e^{\lambda} - 1} \int_0^{\lambda} \sum_{j=1}^{\infty} \frac{u^{j-1}}{j!} \, du$$
$$= \frac{\lambda}{e^{\lambda} - 1} \int_0^{\lambda} \frac{du}{u} \sum_{j=1}^{\infty} \frac{u^j}{j!},$$

which is equal to (27).

(b) When the empirical null is used, the local fdr estimate (18) is $\widehat{\text{fdr}}_k = \hat{y}_k/y_k$, where $\hat{y}_k = N\Delta\hat{p}_0\hat{f}_0(t_k)$. Notice that when evaluated at $t_k \notin S_0$, the numerator $\hat{y}_k$ is independent of the denominator $y_k$. Thus,

$$\text{E}[\widehat{\text{fdr}}_k | y_k > 0] = \text{E}\left(\frac{\hat{y}_k}{y_k}\Big| y_k > 0\right) = \frac{\lambda_{k,0}}{\lambda_k}\frac{\text{E}(\hat{y}_k)}{\lambda_{k,0}}\text{E}\left(\frac{\lambda_k}{y_k}\Big| y_k > 0\right) \approx \text{fdr}_k \, b_k \zeta(\lambda_k),$$

where the vector $\boldsymbol{b}_{\infty}$ is obtained as follows. By (8) and (9), $\boldsymbol{\lambda}_0 = \exp(\boldsymbol{X}\boldsymbol{\eta}^+ + \boldsymbol{h})$ and $\text{E}(\hat{\boldsymbol{y}}) = \text{E}[\exp(\boldsymbol{X}\hat{\boldsymbol{\eta}}^+ + \boldsymbol{h})] \to \exp(\boldsymbol{X}\hat{\boldsymbol{\eta}}_{\infty}^+ + \boldsymbol{h})$ as $N \to \infty$. Dividing entry by entry gives $\boldsymbol{b}_{\infty} = \exp(\boldsymbol{X}(\hat{\boldsymbol{\eta}}_{\infty}^+ - \boldsymbol{\eta}^+))$. □

PROOF OF THEOREM 1. The form of expression (29) follows from Efron (2007a), Lemma 1. A similar argument as in Efron (2007a), Lemma 2, gives that the entries of $\boldsymbol{\delta}$ are

(39) $$\delta_{kl} \approx \int_{-1}^{1} R_{kl}(\rho) \, dG(\rho), \qquad R_{kl}(\rho) = \frac{f_0(t_k, t_l; \rho)}{f_0(t_k)f_0(t_l)} - 1.$$

Replacing (28) in (39) gives (30). □
**Acknowledgments.** The author thanks Bradley Efron and Jonathan Taylor for their guidance, Bob Dougherty for providing the DTI data, and Christoph Lange for providing the FBAT analysis results from the FHS SNP data.

Department of Biostatistics
Harvard School of Public Health
Dana-Farber Cancer Institute
44 Binney Street, CLS-11007
Boston, Massachusetts 02115
USA
E-mail: armins@hsph.harvard.edu